\pdfoutput=1
\documentclass[%
 aip,
 amsmath,amssymb,
twocolumn,
longbibliography,
 reprint,%
]{revtex4-1}

\usepackage{hyperref}
\hypersetup{colorlinks,breaklinks,linkcolor=Blue,urlcolor=Blue,anchorcolor=Blue,citecolor=Blue}
\usepackage{color}
\definecolor{DarkBlue}{rgb}{0.0,0.08,0.45}
\definecolor{Blue}{rgb}{0.0,0.0,1.0}
\definecolor{Red}{rgb}{1.0,0.0,0.0}
\definecolor{RedOrange}{rgb}{0.9,0.0,0.2}
\definecolor{dgreen}{RGB}{0,128,0}
\definecolor{dgray}{gray}{0.3}
\newcommand*{\citen}{}% generate error, if `\citen` is already in use
\DeclareRobustCommand*{\citen}[1]{%
  \begingroup
    \romannumeral-`\x % remove space at the beginning of \setcitestyle
    \setcitestyle{numbers}%
    \cite{#1}%
  \endgroup
}

\newcommand{\nbd}{\nobreakdash}

\usepackage{graphicx}% Include figure files
\usepackage{dcolumn}% Align table columns on decimal point
\usepackage{bm}% bold math
\usepackage[utf8]{inputenc}
\usepackage[T1]{fontenc}
\usepackage{mathptmx}
\usepackage{etoolbox}

\usepackage{multirow}
\usepackage{longtable}
\usepackage{tablefootnote}
\usepackage[table]{xcolor}

\usepackage{comment}

\newcolumntype{.}{D{.}{.}{-1}}
\newcommand{\mc}[3]{\multicolumn{#1}{#2}{#3}}
\newcommand{\tworow}[1]{\multirow{2}{*}{#1}}

\newcommand{\fns}{\footnotesize}

\usepackage{pifont}

\newcommand{\fig}[2]{\scalebox{#1}{\includegraphics{#2}}}

\newcommand{\abws}{BW\nbd-s2($\alpha$)}
\newcommand{\abwsaeq}[1]{BW\nbd-s2($\alpha=#1$)}

\newcommand{\cc}[1]{\cellcolor{#1}}
\definecolor{maxred}{RGB}{231, 114, 111}
\definecolor{mingreen}{RGB}{122, 188, 129}

\definecolor{r1c3}{RGB}{238, 160, 124}
\definecolor{r1c4}{RGB}{244, 197, 135}
\definecolor{r1c5}{RGB}{250, 226, 143}
\definecolor{r1c6}{RGB}{249, 235, 145}
\definecolor{r1c7}{RGB}{160, 202, 133}
\definecolor{r1c8}{RGB}{135, 193, 131}
\definecolor{r1c9}{RGB}{142, 195, 131}
\definecolor{r1c10}{RGB}{252, 236, 146}
\definecolor{r1c11}{RGB}{247, 209, 138}
\definecolor{r2c1}{RGB}{206, 220, 140}
\definecolor{r2c2}{RGB}{228, 227, 142}
\definecolor{r2c4}{RGB}{141, 195, 131}
\definecolor{r2c5}{RGB}{206, 220, 140}
\definecolor{r2c6}{RGB}{252, 236, 146}
\definecolor{r2c7}{RGB}{248, 217, 141}
\definecolor{r2c8}{RGB}{246, 203, 137}
\definecolor{r2c9}{RGB}{243, 190, 133}
\definecolor{r2c10}{RGB}{238, 161, 124}
\definecolor{r3c2}{RGB}{251, 229, 144}
\definecolor{r3c3}{RGB}{244, 194, 134}
\definecolor{r3c4}{RGB}{202, 217, 139}
\definecolor{r3c6}{RGB}{132, 192, 130}
\definecolor{r3c7}{RGB}{154, 200, 133}
\definecolor{r3c8}{RGB}{202, 217, 139}
\definecolor{r3c9}{RGB}{252, 236, 146}
\definecolor{r3c10}{RGB}{246, 204, 137}
\definecolor{r3c11}{RGB}{234, 140, 118}
\definecolor{r4c1}{RGB}{238, 160, 124}
\definecolor{r4c2}{RGB}{129, 190, 129}
\definecolor{r4c3}{RGB}{252, 236, 146}
\definecolor{r4c4}{RGB}{129, 190, 129}
\definecolor{r4c6}{RGB}{153, 199, 133}
\definecolor{r4c7}{RGB}{197, 216, 138}
\definecolor{r4c8}{RGB}{252, 236, 146}
\definecolor{r4c9}{RGB}{248, 216, 140}
\definecolor{r4c10}{RGB}{241, 179, 130}
\definecolor{r5c3}{RGB}{241, 180, 130}
\definecolor{r5c4}{RGB}{251, 231, 144}
\definecolor{r5c5}{RGB}{149, 198, 132}
\definecolor{r5c6}{RGB}{135, 190, 129}
\definecolor{r5c7}{RGB}{156, 200, 133}
\definecolor{r5c8}{RGB}{198, 216, 138}
\definecolor{r5c9}{RGB}{252, 236, 146}
\definecolor{r5c10}{RGB}{247, 210, 138}
\definecolor{r5c11}{RGB}{238, 160, 124}
\definecolor{r6c3}{RGB}{127, 190, 129}
\definecolor{r6c4}{RGB}{156, 200, 133}
\definecolor{r6c5}{RGB}{199, 216, 138}
\definecolor{r6c6}{RGB}{225, 226, 142}
\definecolor{r6c7}{RGB}{250, 235, 145}
\definecolor{r6c8}{RGB}{249, 221, 142}
\definecolor{r6c9}{RGB}{246, 206, 137}
\definecolor{r6c10}{RGB}{240, 176, 129}
\definecolor{r6c11}{RGB}{232, 121, 114}
\definecolor{r7c2}{RGB}{235, 144, 120}
\definecolor{r7c3}{RGB}{139, 194, 131}
\definecolor{r7c4}{RGB}{171, 207, 135}
\definecolor{r7c5}{RGB}{211, 221, 140}
\definecolor{r7c6}{RGB}{231, 228, 143}
\definecolor{r7c7}{RGB}{252, 236, 146}
\definecolor{r7c8}{RGB}{248, 218, 141}
\definecolor{r7c9}{RGB}{244, 200, 136}
\definecolor{r7c10}{RGB}{240, 169, 127}
\definecolor{r8c1}{RGB}{246, 202, 137}
\definecolor{r8c2}{RGB}{240, 173, 128}
\definecolor{r8c3}{RGB}{185, 211, 137}
\definecolor{r8c5}{RGB}{135, 193, 131}
\definecolor{r8c6}{RGB}{164, 203, 134}
\definecolor{r8c7}{RGB}{204, 218, 139}
\definecolor{r8c8}{RGB}{252, 236, 146}
\definecolor{r8c9}{RGB}{249, 219, 141}
\definecolor{r8c10}{RGB}{242, 184, 131}
\definecolor{r9c2}{RGB}{251, 232, 145}
\definecolor{r9c3}{RGB}{241, 178, 130}
\definecolor{r9c4}{RGB}{246, 207, 138}
\definecolor{r9c5}{RGB}{250, 228, 144}
\definecolor{r9c6}{RGB}{252, 236, 146}
\definecolor{r9c7}{RGB}{217, 223, 141}
\definecolor{r9c8}{RGB}{189, 213, 137}
\definecolor{r9c9}{RGB}{166, 205, 134}
\definecolor{r9c10}{RGB}{136, 193, 131}
\definecolor{r10c2}{RGB}{188, 213, 137}
\definecolor{r10c3}{RGB}{238, 162, 125}
\definecolor{r10c4}{RGB}{244, 196, 134}
\definecolor{r10c5}{RGB}{249, 220, 141}
\definecolor{r10c6}{RGB}{245, 223, 140}
\definecolor{r10c7}{RGB}{252, 236, 146}
\definecolor{r10c8}{RGB}{194, 215, 138}
\definecolor{r10c9}{RGB}{155, 200, 133}
\definecolor{r10c11}{RGB}{160, 202, 133}
\definecolor{r11c2}{RGB}{251, 235, 146}
\definecolor{r11c3}{RGB}{238, 160, 124}
\definecolor{r11c4}{RGB}{244, 196, 135}
\definecolor{r11c5}{RGB}{250, 224, 143}
\definecolor{r11c6}{RGB}{252, 236, 146}
\definecolor{r11c7}{RGB}{219, 221, 139}
\definecolor{r11c8}{RGB}{198, 216, 138}
\definecolor{r11c9}{RGB}{178, 209, 136}
\definecolor{r11c10}{RGB}{149, 198, 132}
\definecolor{r12c2}{RGB}{252, 236, 146}
\definecolor{r12c3}{RGB}{237, 156, 123}
\definecolor{r12c4}{RGB}{244, 194, 134}
\definecolor{r12c5}{RGB}{251, 229, 144}
\definecolor{r12c6}{RGB}{219, 224, 141}
\definecolor{r12c7}{RGB}{168, 205, 134}
\definecolor{r12c8}{RGB}{133, 192, 130}
\definecolor{r12c10}{RGB}{169, 206, 135}
\definecolor{r12c11}{RGB}{248, 216, 140}
\definecolor{r13c2}{RGB}{252, 236, 146}
\definecolor{r13c3}{RGB}{237, 154, 123}
\definecolor{r13c4}{RGB}{243, 191, 133}
\definecolor{r13c5}{RGB}{250, 223, 143}
\definecolor{r13c6}{RGB}{241, 231, 144}
\definecolor{r13c7}{RGB}{166, 205, 134}
\definecolor{r13c10}{RGB}{240, 231, 144}
\definecolor{r13c11}{RGB}{244, 197, 135}
\definecolor{r14c2}{RGB}{252, 236, 146}
\definecolor{r14c3}{RGB}{237, 156, 123}
\definecolor{r14c4}{RGB}{244, 193, 134}
\definecolor{r14c5}{RGB}{250, 226, 143}
\definecolor{r14c6}{RGB}{228, 227, 142}
\definecolor{r14c7}{RGB}{161, 202, 133}
\definecolor{r14c10}{RGB}{206, 220, 140}
\definecolor{r14c11}{RGB}{246, 204, 137}

\definecolor{t2r1c1}{RGB}{156, 200, 133}
\definecolor{t2r1c4}{RGB}{141, 195, 131}
\definecolor{t2r1c5}{RGB}{191, 213, 137}
\definecolor{t2r1c6}{RGB}{220, 224, 141}
\definecolor{t2r1c7}{RGB}{252, 236, 146}
\definecolor{t2r1c8}{RGB}{250, 224, 143}
\definecolor{t2r1c9}{RGB}{247, 213, 140}
\definecolor{t2r1c10}{RGB}{243, 191, 133}
\definecolor{t2r1c11}{RGB}{237, 151, 122}
\definecolor{t2r2c3}{RGB}{151, 199, 133}
\definecolor{t2r2c4}{RGB}{166, 205, 134}
\definecolor{t2r2c5}{RGB}{202, 217, 139}
\definecolor{t2r2c6}{RGB}{227, 227, 142}
\definecolor{t2r2c7}{RGB}{252, 236, 146}
\definecolor{t2r2c8}{RGB}{251, 229, 144}
\definecolor{t2r2c9}{RGB}{241, 179, 130}
\definecolor{t2r2c10}{RGB}{247, 214, 140}
\definecolor{t2r2c11}{RGB}{242, 184, 131}

\makeatletter
\def\@email#1#2{%
 \endgroup
 \patchcmd{\titleblock@produce}
  {\frontmatter@RRAPformat}
  {\frontmatter@RRAPformat{\produce@RRAP{*#1\href{mailto:#2}{#2}}}\frontmatter@RRAPformat}
  {}{}
}%
\makeatother

\usepackage{xr-hyper}
\makeatletter
\newcommand*{\addFileDependency}[1]{% argument=file name and extension
  \typeout{(#1)}
  \@addtofilelist{#1}
  \IfFileExists{#1}{}{\typeout{No file #1.}}
}
\makeatother

\addFileDependency{supportingInformation.tex}
\addFileDependency{supportingInformation.aux}

\begin{document}

\preprint{AIP/123-QED}

\title{
Optimizing the Regularization in Size-Consistent Second-Order Brillouin-Wigner Perturbation Theory
}
\author{Kevin Carter-Fenk}
\affiliation{
Kenneth S. Pitzer Center for Theoretical Chemistry, Department of Chemistry, University of California, Berkeley, CA 94720, USA.
}
\author{James Shee}
\affiliation{ 
Kenneth S. Pitzer Center for Theoretical Chemistry, Department of Chemistry, University of California, Berkeley, CA 94720, USA.
}
\affiliation{{Department of Chemistry, Rice University, Houston, TX 77005, USA}}
\author{Martin Head-Gordon}%
 \email{mhg@cchem.berkeley.edu}
 \affiliation{ 
Kenneth S. Pitzer Center for Theoretical Chemistry, Department of Chemistry, University of California, Berkeley, CA 94720, USA.
}
\affiliation{
Chemical Sciences Division, Lawrence Berkeley National Laboratory, Berkeley, CA 94720, USA
}

\date{\today}% It is always \today, today,

\begin{abstract}
Despite its simplicity and relatively low computational cost, second-order M{\o}ller-Plesset perturbation theory (MP2) is well-known to overbind noncovalent interactions between polarizable monomers and some organometallic bonds.  In such situations, the pairwise-additive correlation energy expression in MP2 is inadequate.  Although energy-gap dependent amplitude regularization can substantially improve the accuracy of conventional MP2 in these regimes, the same regularization parameter worsens the accuracy for small molecule thermochemistry and density-dependent properties.  Recently,
we proposed a repartitioning of Brillouin-Wigner perturbation theory that is size-consistent to second order (BW-s2), and a free parameter ($\alpha$)  was set to recover the exact dissociation limit of H$_2$ in a minimal basis set. Alternatively $\alpha$ can be viewed as a regularization parameter, where each value of $\alpha$ represents a valid variant of BW-s2, which we denote as BW-s2($\alpha$). In this work, we semi-empirically optimize $\alpha$ for noncovalent interactions, thermochemistry, alkane conformational energies, electronic response properties,
and transition metal datasets, leading to improvements in accuracy relative to
the {\em ab initio}\ parameterization of BW-s2 and MP2.
We demonstrate that the optimal $\alpha$ parameter ($\alpha=4$) is more transferable across chemical
problems than energy-gap-dependent regularization parameters.
This is attributable to the fact that the \abws\ regularization strength depends on all of the information
encoded in the $\mathbf{t}$ amplitudes rather than just orbital energy differences.
While the computational scaling of \abws\ is iterative $\mathcal{O}(N^5)$, this effective and transferable approach to amplitude regularization is a promising route to incorporate higher-order correlation effects at second-order cost.
\end{abstract}

\maketitle

%%%%%%%%
% FOR JCP COMMUNICATION: 3500 word limit, not including tables/figures
%%%%%%%%%

% NOTE: In the spirit of a JCP communication, and to my dismay, I might be inclined to remove large swaths of this (nicely written) introduction.
% I will keep large sections hanging out in comment blocks in case we choose to submit this elsewhere.

%\section{\label{sec:Intro} Introduction}
M{\o}ller-Plesset perturbation theory (MP2) is a remarkable theoretical model chemistry with a simple pairwise additive form of the electron correlation energy and relatively low $\mathcal{O}(N^5)$ compute cost scaling.  It is the simplest \emph{ab initio} theory that can approximately describe many forms of weak electron correlations, most notably dispersion but also short-ranged exchange effects.
The correlation energy in the canonical molecular orbital basis can be written
\begin{equation}\label{eq:MP2Energy}
	E_c = -\frac{1}{4}\sum\limits_{ijab}\frac{|\mathbb{I}_{ijab}|^2}{\varepsilon_a+\varepsilon_b-\varepsilon_i-\varepsilon_j} = -\frac{1}{4}\sum\limits_{ijab}\frac{|\mathbb{I}_{ijab}|^2}{\Delta_{ij}^{ab}} \; ,
\end{equation}
where $\mathbb{I}_{ijab} = (ij||ab)$ 
are antisymmetrized two-electron integrals and $\varepsilon_p$ is
the $p$-th orbital eigenvalue.
Throughout this text we apply the standard notation where
\{$i, j, k$\dots\} refer to occupied orbitals,
\{$a, b, c$\dots\} to unoccupied orbitals,
and \{$p, q, r$\dots\} to arbitrary (occupied or virtual) orbitals.

Formally, MP2 has many desirable properties.
For example, it is free of delocalization errors, unlike the widely popular
density functional theory (DFT).\cite{ZhaYan98b,MorCohYan06,MorCohYan08}
In contrast to both DFT and the direct random phase approximation (RPA),\cite{CheVooAge17} there is no self-correlation error.
Consistent with Pople's high standards for an approximate model chemistry\cite{Pop99} (at the heart of which are formal principles which are practically useful in chemical predictions), MP2 is size-consistent, size-extensive,
and orbital invariant.\cite{SzaOst77}
A model is size-consistent if the total energy of a
supersystem comprised of noninteracting subsystems
is the same as the sum of the energies of the isolated subsystems; this is an essential property when studying
phenomena such as bond breaking.
Second, a method is size-extensive if the total correlation energy in a linear chain of atoms
grows linearly with number of electrons,
which is essential for reaching the thermodynamic limit.
Third, a method that yields the same correlation energy despite arbitrary orbital rotations in the occupied (or virtual) subspace is considered to be orbital invariant -- a property that enables transformations to chemically-relevant bases such as the natural orbital or localized orbital representations.
%We note that, unlike Full Configuration Interaction,
%MP2 (along with coupled cluster methods) is not variational,
%{\em i.e.}\ it is possible to attain an energy below the exact energy. 

MP2 routinely outperforms
Hartree-Fock (HF)
theory across myriad test sets with respect to
experimental or near-exact numerical reference values.\cite{SheRoiLet21}
The accuracy of MP2 can be very high in the case of closed-shell and small organic molecules, and can
exceed the accuracy of popular DFT functionals for
important chemical properties such as reaction
barrier heights (which are sensitive to delocalization errors).\cite{GonXavLi19}
Indeed, MP2 is the most popular wavefunction component to be
incorporated into double-hybrid density functionals, with
promising results in many chemically-relevant situations.\cite{wB97X-2,ShaTouSav11,SanAda13,GoeGri14,BreCioSan16,wB97M2,KalTou18,MarSan20}

However, over the years many shortcomings of MP2 have been found.  It is well-known that perturbation theory in general is not suitable for multi-reference states, in which higher order (connected) excitations are required for a qualitatively correct description of the wavefunction.
%It is well-known today that MP2-type methods should simply not be used in these regimes. 
In addition, MP2 (and even higher orders of perturbation theory) can fail in certain cases where the reference determinant is severely spin-contaminated.\cite{HubCar80,MurDav91,LauStaGau91,AmoAndNic91,KnoAndAmo91,LeeJay93}
In strongly correlated cases, this is a fatal issue; however, in weakly correlated systems, where the spin-symmetry breaking is artificial (i.e. due to deficiencies in the model chemistry's treatment of dynamic correlation), MP2 with a restricted open-shell (RO) reference determinant can at times remedy this situation.\cite{KnoAndAmo91}
%However, for, {\em e.g.}, transition metal complexes of high spin multiplicity ROHF calculations are challenging (or impossible) to converge.  
Approximate Brueckner orbital approaches,\cite{LocHea07,AkiKawTen11}
{\em e.g.}\ orbital optimized (MP2) methods,\cite{NeeSchKos09,StuHea13,ShaStuSun15,SoyBoz15,RazStuHea17,LeeHea18,BozUnaAla20,BehFin22}
can also clean up spin-symmetry breaking at the level of the Hartree-Fock orbitals.

Yet even after putting the above issues (stemming from multireference character and open-shell situations) aside, there are still serious difficulties that have historically limited the use and accuracy of MP2 approaches.  For example, when
bonds are stretched, the denominator of Eq.~\ref{eq:MP2Energy} can become zero, causing the correlation energy to diverge.
This severely complicates the calculation of smooth potential energy surfaces.  Interestingly, pair energies (corresponding to occupied orbitals $i$ and $j$) can be overestimated even at equilibrium geometries, most notably in the cases of dispersion-dominated noncovalent interaction energies (NC) among polarizable monomers (e.g., those with conjugated $\pi$ systems)\cite{SinValShe02,SinShe04b,NguCheAge20}
and organometallic bonds involving, e.g., metal-carbonyl moieties.\cite{NeeSchKos09,DohHanSte18,SheLoiHai21}
%A number of physically-motivated regularization schemes %-- both empirical and formally-derivable --
%have been suggested and assessed,
%with energy gap-dependent regularization among the most promising.
Physically-motivated regularization schemes that aim to remove divergences due to the energy denominator in Eq.~\ref{eq:MP2Energy} offer a promising approach to ameliorating the above problems.
One example, $\kappa$-MP2, takes the form,
\begin{equation}
	E_c = -\frac{1}{4}\sum\limits_{ijab} \frac{|\mathbb{I}_{ijab}|^2}{\Delta_{ij}^{ab}}
	\Big(1 - e^{-\kappa\Delta_{ij}^{ab}}\Big)^2 \; ,
\end{equation}
and improves upon conventional MP2 for large-molecule
NC and closed-shell
transition-metal thermochemistry (TMTC) by factors of 5 and 2, respectively.\cite{SheRoiLet21} 

%The energy-gap dependent protocols for MP2 regularization, however, lack the desired level of transferrability to be widely used in a black-box fashion.
Despite such notable improvements over conventional MP2, energy-gap dependent protocols for MP2 regularization lack the
desired level of transferability required to be widely used in a black-box fashion.
For example, with regularization parameters optimized for NC and transition metal systems, the accuracy for main-group thermochemistry (TC)
and electronic response properties is notably deteriorated -- at times these regularized MP2 approaches are worse than conventional MP2 by factors of 2 or 3.\cite{SheRoiLet21} Similarly, $\kappa$-MP2 demonstrated very promising improvements in accuracy relative to MP2 for NMR chemical shifts only when  element-specific $\kappa$ values were employed.\cite{WonGanLiu23}  The prospect of developing a more transferable approach to regularized second-order perturbation theory, which preserves high accuracy for NC and TM datasets, is the primary motivation for the present work.

Brillouin-Wigner perturbation theory (BWPT)\cite{Len30,Bri32,Wig35b,HubWil10} is an alternative to the Rayleigh-Schr\"odinger approach (the latter gives rise to MP2).  We recently proposed a size-consistent variant of second-order BWPT, which naturally regularizes the $\mathbf{t}$-amplitudes by shifting the occupied orbital energies in the denominator to lower values, thus increasing the effective orbital energy gaps and damping artificially overestimated amplitudes.\cite{CarHea23}
 The single free parameter in our BW-s2 approach was determined such that the dissociation limit
 of a system with two electrons in two orbitals (e.g., the H$_2$ molecule in a minimal basis set) is exact.
 Importantly, this model, which we refer to as BW-s2, is size-consistent, size-extensive, and orbital invariant.
 While BW-s2
 was found to be less accurate than (optimally parameterized) $\kappa$-MP2
 in cases where exceptionally strong regularization was required,
 for a wide variety of main group TC its performance is superior to $\kappa$-MP2
 and conventional MP2.
 In this work we aim to explore the landscape of the free parameter, which we will call $\alpha$,
 by investigating many different data sets representative of NC,
 large-gap TMTC, main-group TC,
 barrier heights, and molecular dipoles and polarizabilities.

Recently, we have shown that with the following repartitioning of the Hamiltonian (with $\hat{R}$, a one-electron regularizer operator):
\begin{equation}
	\hat{H} = \hat{\bar{H}}_0 + \lambda\hat{\bar{V}} \; ,
\end{equation}
where
\begin{equation}\label{eq:ShiftedH0}
	\begin{split}
	\hat{\bar{H}}_0 &= \hat{H}_0 + \hat{R}\\
	\hat{\bar{V}} &= \hat{V} - \hat{R}
	\end{split}
\end{equation}
the second-order BWPT correction becomes
\begin{equation}\label{eq:PT2Energy}
E^{(2)} = \sum\limits_{k\neq0}\frac{\langle\Phi_0|\hat{\bar{V}}|\Phi_k\rangle
	\langle\Phi_k|\hat{\bar{V}}|\Phi_0\rangle}{(\bar{E}_0-\bar{E}_k)+E^{(2)}} \; ,
\end{equation}
where $\bar{E}_0$ and $\bar{E}_k$ are eigenvalues
of the shifted zero-order Hamiltonian, $\hat{\bar{H}}_0$.
The energy gap $\bar{\Delta} = \bar{E}_0-\bar{E}_k$
generally satisfies the relationship
$\bar{\Delta}\geq\Delta$ where $\Delta = E_0 - E_k$
is the gap derived from the usual eigenvalues of
the unshifted $\hat{H}_0$. The use of these barred
quantities is the only difference between our approach and
typical BWPT.

%In the tensor formulation of many-body perturbation
%theory,\cite{HeaMasWhi98,LeeMasHea00,DiSJunHea05} which by construction guarantees the desired property of orbital invariance, Eq.~\ref{eq:ShiftedH0} leads to
The above expressions are general, and while
there are infinitely many partitions of $\hat{\bar{H}}_0$,
there are finitely many that are size-consistent.
We have chosen a particular form of $\hat{R}$ that is size-consistent,
and is represented in an arbitrary molecular orbital basis as,
\begin{equation}\label{eq:Rtensor}
	R_{ijkl}^{abcd} = \frac{\alpha}{2}(W_{ik}\delta_{jl} + \delta_{ik}W_{jl})\delta_{ac}\delta_{bd} \;.
\end{equation}
with,
\begin{equation}\label{eq:Wmatrix}
	W_{ij} = \frac{1}{2}\sum\limits_{kab} \left[ t_{ik}^{ab}(jk||ab) + t_{jk}^{ab}(ik||ab) \right] \;.
\end{equation}
The generalized tensor formulation of the second-order amplitude equation
then reads,\cite{HeaMasWhi98,LeeMasHea00,DiSJunHea05}
\begin{equation}\label{eq:MFCtensor}
\sum\limits_{klcd}\big(\Delta_{ijkl}^{abcd} + R_{ijkl}^{abcd}\big)\cdot t_{kl}^{cd} = -\mathbb{I}_{ijab}.
\end{equation}
where 
\begin{equation}\label{eq:DeltaTensor}
	\Delta_{ijkl}^{abcd} = (F_{ac}\delta_{bd} + \delta_{ac}F_{bd})\delta_{ik}\delta_{jl}
	- (F_{ik}\delta_{jl} + \delta_{ik}F_{jl})\delta_{ac}\delta_{bd} \; .
\end{equation}

\begin{table*}[!!ht]
\centering
\setlength{\tabcolsep}{4pt}
\caption{
	Root-mean-square error in kcal\slash mol across chemical benchmark sets.
}\label{tbl:Data}
\begin{tabular}{ll ... .. . .. .. ..}
\hline\hline
\tworow{Type} &
\mc{1}{c}{\tworow{Benchmark}} & \mc{1}{c}{\multirow{2}{*}{HF}} & \mc{1}{c}{\multirow{2}{*}{MP2}} &
\mc{1}{c}{\tworow{$\kappa$-MP2$^a$}} & \mc{8}{c}{\abws: $\alpha$-parameter}\\ \cline{6-14}
& & & & & \mc{1}{c}{1.0} & \mc{1}{c}{2.0} & \mc{1}{c}{3.0} & \mc{1}{c}{3.5} & \mc{1}{c}{4.0} & \mc{1}{c}{4.5} & \mc{1}{c}{5.0} & \mc{1}{c}{6.0} & \mc{1}{c}{8.0}\\
\hline
& L7$^b$ & 27.62 & \cc{maxred}9.49 & \cc{mingreen}1.25	&  \cc{r1c3}6.55	& \cc{r1c4}4.30 &	\cc{r1c5}2.57	& \cc{r1c6}1.93 &	\cc{r1c7}1.47 &	\cc{r1c8}1.33 &	\cc{r1c9}1.37
& \cc{r1c10}1.94 & \cc{r1c11}3.57\\
& A24 & 1.64 & \cc{r2c1}0.14 & \cc{r2c2}0.15 &	\cc{mingreen}0.10 &	\cc{r2c4}0.11 &	\cc{r2c5}0.14 & \cc{r2c6}0.16	& \cc{r2c7}0.19 &	\cc{r2c8}0.21 &	\cc{r2c9}0.23
& \cc{r2c10}0.27 & \cc{maxred}0.35\\
NC$^c$ & X31 & 3.43 & \cc{maxred}0.69 & \cc{r3c2}0.38 &	\cc{r3c3}0.47 &	\cc{r3c4}0.32 &	\cc{mingreen}0.25 &	\cc{r3c6}0.26 &	\cc{r3c7}0.28 &	\cc{r3c8}0.32 &	\cc{r3c9}0.36
& \cc{r3c10}0.45 & \cc{r3c11}0.62\\
& S66 & 4.70 & \cc{r4c1}0.75 & \cc{r4c2}0.35 &	\cc{r4c3}0.49 &	\cc{r4c4}0.35 &	\cc{mingreen}0.34 &	\cc{r4c6}0.38 &	\cc{r4c7}0.43 &	\cc{r4c8}0.49 &	\cc{r4c9}0.56 &
\cc{r4c10}0.68 & \cc{maxred}0.91\\
& S22$^d$ & 6.18 & \cc{maxred}1.36 & \cc{mingreen}0.35 &	\cc{r5c3}0.91 &	\cc{r5c4}0.58 &	\cc{r5c5}0.39 &	\cc{r5c6}0.37 &	\cc{r5c7}0.40 &	\cc{r5c8}0.47
& \cc{r5c9}0.55
& \cc{r5c10}0.72 & \cc{r5c11}1.05\\
\hline
   & HTBH38 & 19.02 & \cc{mingreen}5.03 & \cc{maxred}6.74 & \cc{r6c3}5.06 &	\cc{r6c4}5.21 &	\cc{r6c5}5.42 &	\cc{r6c6}5.54 &	\cc{r6c7}5.66 &	\cc{r6c8}5.79 &	\cc{r6c9}5.92
   & \cc{r6c10}6.18 & \cc{r6c11}6.68\\
TC$^e$ & NHTBH38 & 16.29 & \cc{mingreen}2.40 & \cc{r7c2}5.61 & \cc{r7c3}2.68 &	\cc{r7c4}3.16 &	\cc{r7c5}3.73 &	\cc{r7c6}4.01 &	\cc{r7c7}4.28 &	\cc{r7c8}4.54 &	\cc{r7c9}4.79 & \cc{r7c10}5.25 & \cc{maxred}6.06\\
   & W4-11 & 55.61 & \cc{r8c1}7.55 & \cc{r8c2}8.28 & \cc{r8c3}6.18 &	\cc{mingreen}5.63 &	\cc{r8c5}5.76 &	\cc{r8c6}6.01 &	\cc{r8c7}6.33 &	\cc{r8c8}6.71 &	\cc{r8c9}7.13
   & \cc{r8c10}8.01 & \cc{maxred}9.79\\
\hline
& MOR39 & 22.15  & \cc{maxred}14.13 & \cc{r9c2}6.49 & \cc{r9c3}9.93 &	\cc{r9c4}8.11 &	\cc{r9c5}6.77 &	\cc{r9c6}6.23 &	\cc{r9c7}5.78 &	\cc{r9c8}5.40 &	\cc{r9c9}5.09
& \cc{r9c10}4.66 & \cc{mingreen}4.43\\
TMTC$^f$ & MC09  & 29.34 & \cc{maxred}14.42 & \cc{r10c2}6.18 & \cc{r10c3}11.26 &	\cc{r10c4}9.08 &	\cc{r10c5}7.56 &	\cc{r10c6}6.99 &	\cc{r10c7}6.55 &	\cc{r10c8}6.22 &	\cc{r10c9}5.98 &
\cc{mingreen}5.76 & \cc{r10c11}6.01\\
& AuIrPt13 & 6.40 & \cc{maxred}4.30 & \cc{r11c2}2.60 &  \cc{r11c3}3.64 &	\cc{r11c4}3.13 &	\cc{r11c5}2.74 &	\cc{r11c6}2.58 &	\cc{r11c7}2.45 &	\cc{r11c8}2.33 &	\cc{r11c9}2.23 &
\cc{r11c10}2.08 & \cc{mingreen}1.93\\
\hline
& ACONF12 & 4.18 & \cc{maxred}0.83 & \cc{r12c2}0.26 & \cc{r12c3}0.63 &	\cc{r12c4}0.45 &	\cc{r12c5}0.29 &	\cc{r12c6}0.22 &	\cc{r12c7}0.16 &	\cc{r12c8}0.11 &	\cc{mingreen}0.10 &
\cc{r12c10}0.16 & \cc{r12c11}0.36\\
ACONFL$^g$ & ACONF16 & 4.11 & \cc{maxred}0.85 & \cc{r13c2}0.20 & \cc{r13c3}0.63 &	\cc{r13c4}0.44 &	\cc{r13c5}0.27 &	\cc{r13c6}0.19 &	\cc{r13c7}0.13 &	\cc{mingreen}0.09 &	\cc{mingreen}0.09 &
\cc{r13c10}0.19 & \cc{r13c11}0.40\\
& ACONF20 & 5.00 & \cc{maxred}1.07 & \cc{r14c2}0.29 & \cc{r14c3}0.80  & \cc{r14c4}0.56 &	\cc{r14c5}0.35 &	\cc{r14c6}0.27 &	\cc{r14c7}0.19 &	\cc{mingreen}0.15 &	\cc{mingreen}0.15 &
\cc{r14c10}0.24 & \cc{r14c11}0.49\\
\hline\hline
\mc{14}{l}{\fns
    $^a$$\kappa=1.1$
}\\
\mc{14}{l}{\fns
    $^b$heavy-aug-cc-pVDZ\slash heavy-aug-cc-pVTZ
    extrapolation to the complete basis set limit
}\\
\mc{14}{l}{\fns
    $^c$Noncovalent interaction energies, aug-cc-pVDZ\slash aug-cc-pVTZ
    extrapolation to the complete basis set limit, unless otherwise noted
}\\
\mc{14}{l}{\fns
    $^d$Using the reference
    data from Ref.~\citen{MarBurShe11}.
}\\
\mc{14}{l}{\fns
    $^e$Thermochemistry, aug-cc-pVTZ\slash aug-cc-pVQZ
    extrapolation to the complete basis set limit
}\\
\mc{14}{l}{\fns
    $^f$Transition metal thermochemistry, MOR39: Def2-TZVPP\slash Def2-ECP,
    MC09: Def2-QZVPP\slash Def2-ECP,
    AuIrPt13: cc-pVTZ\slash cc-pVTZ-PP
}\\
\mc{14}{l}{\fns
    $^g$Alkane conformational isomer energies,
    heavy-aug-cc-pVDZ\slash heavy-aug-cc-pVTZ extrapolation
    to the complete basis set limit
}\\
\end{tabular}
\end{table*}

Given the definition in Eq. \ref{eq:Rtensor}, Eq.~\ref{eq:MFCtensor} (which is orbital invariant) can be solved by rotating the occupied subspace from the canonical basis
into a basis where the matrix $\mathbf{F}_{\text{oo}}+\frac{\alpha}{2}\mathbf{W}$
is diagonal.
To do this, we solve the following eigenvalue equation,
\begin{equation}\label{eq:EigenProblem}
	\bigg(\mathbf{F}_{\text{oo}}+\frac{\alpha}{2}\mathbf{W}\bigg)\mathbf{U}
	= \tilde{\varepsilon}\mathbf{U}
\end{equation}
to obtain a set of dressed occupied orbital eigenvalues.
In this dressed-orbital basis, Eq.~\ref{eq:MFCtensor} can be written as,
\begin{equation}\label{eq:DressedBasisAmplitudes}
	(\varepsilon_a + \varepsilon_b - \tilde{\varepsilon}_i - \tilde{\varepsilon}_j)\tilde{t}_{ij}^{ab}
	= -\tilde{\mathbb{I}}_{ijab}
\end{equation}
leading to, 
\begin{equation}
	\tilde{t}_{ij}^{ab}
	= -\frac{\tilde{\mathbb{I}}_{ijab}}{(\varepsilon_a + \varepsilon_b - \tilde{\varepsilon}_i - \tilde{\varepsilon}_j)}
\end{equation}
and,
\begin{equation}\label{eq:MP2like}
	\tilde{E}_c
	= -\frac{1}{4}\sum\limits_{ijab}\frac{|\tilde{\mathbb{I}}_{ijab}|^2}{(\varepsilon_a + \varepsilon_b - \tilde{\varepsilon}_i - \tilde{\varepsilon}_j)}
\end{equation}
Thus, the dressed eigenvalues $\tilde{\varepsilon}_p$ have the effect of augmenting the original denominator, $\Delta_{ij}^{ab}$, by
adding a correlation contribution to the occupied orbital energies. The undetermined parameter $\alpha$ was set to 1 based on making the theory exact for the 2 electron in 2 orbital problem. Similar ideas have recently been presented in Green's function based perturbation theories,\cite{LanKanZgi16,NeuBaeZgi17,CovTew23}
but unlike these methods, our BW-s2 approach
retains the crucial property of orbital invariance.
There are also notable similarities
between BW-s2
and the perturbation-adapted perturbation theory (PAPT) of Knowles, which
seeks to optimize the partitioning
of $\hat{H}$.\cite{Kno22} Whereas PAPT
costs ${\mathcal O}(N^6)$ already
at second order, BW-s2 scales
much more favorably at
iterative ${\mathcal O}(N^5)$.

In our original set of benchmarks,\cite{CarHea23}
we found that BW-s2 consistently outperforms MP2
across myriad chemical problems, which is very encouraging. However, it was evident that specific, optimal choices of $\kappa$ in $\kappa$-MP2 could significantly outperform BW-s2 in problems where strong regularization was required (such as transition metal thermochemistry).
How much improvement is possible if we lift the restriction of $\alpha=1$, and instead view $\alpha$ as a parameter that controls regularization strength?
That is the question that we will investigate here.

In this work, we benchmark the performance of various values of $\alpha$
against a variety of data sets in an effort
to tune the accuracy of BW-s2 [henceforth, the empirical
variant will be referred to as \abws].
Notably, the particular value of $\alpha$ does
not influence the size-consistency of the method,
but it may be a determining factor in the overall quality
of the results.
We will assess the transferability of the $\alpha$ parameter
across various chemical problems, and
attempt to make a recommendation for a broadly applicable $\alpha$ value.

\begin{table*}[!!ht]
\caption{
	Root-mean-square relative error in \% for electronic properties.
}\label{tbl:Properties}
\begin{tabular}{l ... .. .. .. .. .}
\hline\hline
\mc{1}{c}{\tworow{Benchmark}} & \mc{1}{c}{\tworow{HF$^a$}} & \mc{1}{c}{\tworow{MP2$^b$}} &
\mc{1}{c}{\tworow{$\kappa$-MP2$^{b,c}$}} 
& \mc{8}{c}{\abws$^b$: $\alpha$-parameter}\\ \cline{5-13}
& & & & \mc{1}{c}{1.0} & \mc{1}{c}{2.0} & \mc{1}{c}{3.0} & \mc{1}{c}{3.5} & \mc{1}{c}{4.0} & \mc{1}{c}{4.5} & \mc{1}{c}{5.0} & \mc{1}{c}{6.0} & \mc{1}{c}{8.0}\\
\hline
Dipoles$^d$ & 12.69 & \cc{t2r1c1}3.60 & \cc{maxred}6.81 & \cc{mingreen}3.30 &	\cc{t2r1c4}3.48 &	\cc{t2r1c5}3.88 &	\cc{t2r1c6}4.11 &	\cc{t2r1c7}4.34 &	\cc{t2r1c8}4.58 &	\cc{t2r1c9}4.81
& \cc{t2r1c10}5.25 & \cc{t2r1c11}6.03 \\
Polarizabilities$^e$ & 6.93 & \cc{mingreen}2.16 & \cc{maxred}5.52 & \cc{t2r2c3}2.33 &	\cc{t2r2c4}2.42 &	\cc{t2r2c5}2.61 &	\cc{t2r2c6}2.74 &	\cc{t2r2c7}2.87 &	\cc{t2r2c8}3.02 &	\cc{t2r2c9}3.24
& \cc{t2r2c10}3.34 & \cc{t2r2c11}3.98\\
\hline\hline
\mc{13}{l}{\fns
    $^a$aug-cc-pCVQZ basis set
}\\
\mc{13}{l}{\fns
    $^b$aug-cc-pCVTZ\slash aug-cc-pCVQZ extrapolation to complete basis set limit
}\\
\mc{13}{l}{\fns
    $^c$$\kappa=1.1$
}\\
\mc{13}{l}{\fns
    $^d$Dipole benchmark data from Ref.~\citen{HaiHea18a}
}\\
\mc{13}{l}{\fns
    $^e$Polarizability benchmark data from Ref.~\citen{HaiHea18}
}\\
\end{tabular}
\end{table*}

%\section{Results and Discussion}
The results for all benchmark sets apart from electronic properties
are shown in Table~\ref{tbl:Data}
and are plotted individually as a function of $\alpha$
for each data set in Figures~\ref{fig:NCI_alphas}--\ref{fig:ACONFL_alphas}.
These data include NC for sets of
small dimers such as A24,\cite{A24} S22,\cite{S22} S66,\cite{S66} and the non-I-containing subset
of X40 (hereafter referred to as X31),\cite{X40}
along with the large $\pi$-stacked dimers of L7.\cite{L7}
TC is assessed on H-atom transfer (HTBH38)
and non-H-atom transfer (NHTBH38) sets,\cite{ZhaGonTru05,ZheZhaTru07}
along with the more comprehensive single-reference subset of W4-11.\cite{W4-11}
As compared to our original work, we extend our coverage of
TMTC with
reaction energies from MOR39\cite{DohHanSte18} (a subset of MOR41 with triple-$\zeta$ reference values),\cite{CarHea23}
MC09,\cite{MCO9}
and a set of 13 Au, Pt, and Ir reaction energies
that we call AuPtIr13.\cite{AuIrPt13}
Finally, we also include the ACONFL set of
relative alkane conformational isomer energies.\cite{EhlGriHan22}

The MP2 results always improve on those obtained from HF,
and gap-regularized $\kappa$-MP2 improves further on these
results in all benchmark sets, apart from TC
where the results degrade by up to 3.2~kcal\slash mol.
A similar trend emerges for \abws, with noticeable improvements
over MP2 for NC, TMTC, and ACONFL data sets,
but the results for barrier heights degrade by only about half
as much as $\kappa$-MP2
(0.03--1.9~kcal\slash mol less accurate for
modest parameters in the range $1\leq\alpha\leq4$). 
Furthermore, \abws\ performs roughly 1~kcal\slash mol
better than MP2 on the W4-11 benchmark set
regardless of the particular value of $\alpha$,
whereas $\kappa$-MP2 performs slightly (0.7~kcal\slash mol) worse.
The improvements in NC, TMTC, and ACONFL sets
with minimal degradation in the results for TC
suggest that the \abws\ $\alpha$-parameter is
more transferable than the $\kappa$ in $\kappa$-MP2.\cite{SheRoiLet21,CarHea23}

Regarding the transferability argument, it is
instructive to consider electronic properties
such as dipole moments and polarizabilities
that are shown in Table~\ref{tbl:Properties}.
Whereas $\kappa$-MP2 doubles the errors relative
to MP2 for both dipoles and polarizabilities,
\abws\ exhibits an exceptional flatness in the
errors as a function of $\alpha$, only increasing sharply
at the most severe $\alpha=5.0$ where the errors
nonetheless remain lower than $\kappa$-MP2.
Reference~\citen{SheRoiLet21} reports $\kappa$-MP2 errors
($1.6\geq\kappa\geq1$) for
dipoles that span the
range 4.7--7.5\%
while polarizability errors span 4.2--5.9\%.
Not only are these close to the largest errors that we report for \abws,
but \abwsaeq{1}\ actually improves the results
for dipole moments relative to MP2, whereas $\kappa$-MP2
errors monotonically increase as $\kappa$ decreases.
These results for electronic properties suggest that
the $\alpha$ parameter in \abws\ is indeed much more
transferable between classes of chemical problem
than gap-dependent regularizers.

\begin{figure}
\centering
\fig{1.0}{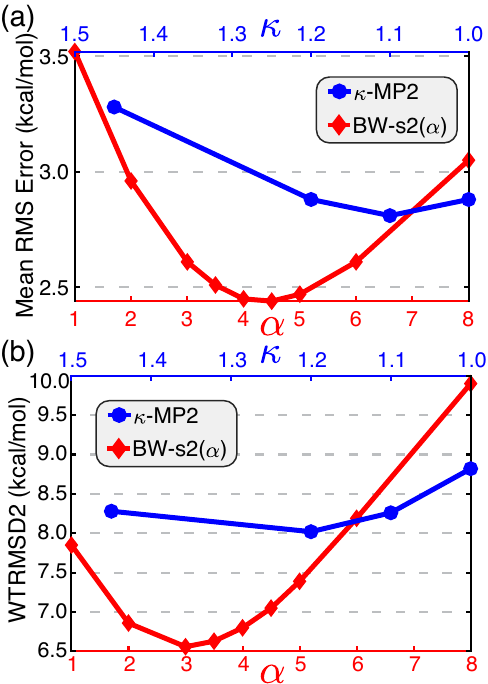}
\caption{
A comparison of errors across
all of the data sets in
Table~\ref{tbl:Data} using (a) the
mean root-mean square deviation (MRMSD)
and (b) the weighted total root-mean square
deviation -- type 2 (WTRMSD2)
for the most successful
range of $\kappa$-MP2
$\kappa$ values
from Ref.~\citen{SheRoiLet21}
contrasted with
BW-s2($\alpha$) $\alpha$ values
from this work.
X-axes are oriented in the direction of increasing 
regularization strength.}\label{fig:RMSD_Metrics}
\end{figure}

Comparison of mean RMSD (MRMSD)
values
in Fig.~\ref{fig:RMSD_Metrics}a
emphasizes the enhanced transferability
of \abws\ relative to $\kappa$-MP2.
The MRMSD, evaluated as a simple average
over RMSD values of each data set
in Table~\ref{tbl:Data},
reveals that \abws\ outperforms
the previously suggested\cite{SheRoiLet21}
optimal $\kappa$-MP2($\kappa=1.1$)
over the wide range of $3\leq\alpha\leq6$.
While this data weighs the error of each
data set on equal footing,
we also report the weighted total RMSD
(WTRMSD2) statistics in Fig.~\ref{fig:RMSD_Metrics}b,
which accounts for the different sizes
and energy scales of each data set.
Specifically, WTRMSD2
is analogous to the type-2 weighted total
mean absolute deviation metric
proposed in Ref.~\citen{GoeHanBau17},
and it is calculated as,
\begin{equation}
    \text{WTRMSD2} = \frac{78.29}{\sum_iN_i}
    \sum\limits_i N_i \frac{\text{RMSD}_i}{|\bar{E}_i|} 
\end{equation}
where $N_i$ is the number of values in set $i$,
$|\bar{E}_i|$ is the average
absolute value of the benchmark
energies in set $i$,
and the constant 78.29~kcal\slash mol was determined
as the average of all $\bar{E}$ values
for all sets.
The WTRMSD2 data reinforce the idea that
\abws\ is more flexible than $\kappa$-MP2,
with an even wider range of $1\leq\alpha\leq5$ that outperform
$\kappa$-MP2($\kappa=1.1$).
Notably, WTRMSD2 suggests
that even the original \abwsaeq{1}
outperforms the best $\kappa$ value.
WTRMSD2 is likely to skew the
results towards nominal performance on TC
properties due to the relative
enormity of the W4-11
set (which contains 745 reactions),
hence the preference for lower $\alpha$
values in this case.
While the optimal value
of $\alpha$ shifts depending on the particular
averaging scheme used,
a value of $\alpha=4$ is roughly
optimal relative to both MRMSD
and WTRMSD2 metrics,
and is likely a sensible compromise
value that performs well for most
chemical problems.

Some particularly interesting highlights are that
\abws\ can reduce errors relative to MP2 in the L7 data set from
9.5~kcal\slash mol to 1.3~kcal\slash mol.
TMTC data can also be improved by a factor of 2--3
relative to the MP2 results, reducing errors
from 14~kcal\slash mol to 4--6~kcal\slash mol
for MOR39 and MCO9 sets if
moderate to large $\alpha$ parameters are applied.
Finally, errors in alkane conformational energies
can be reduced from $\sim1$~kcal\slash mol with MP2
to just 0.1~kcal\slash mol with \abws, achieving
something close to chemical accuracy.
Of course, excellent performance for particular kinds of
chemical problem does not suggest a ``universal''
$\alpha$ value, and there is likely no $\alpha$
parameter that is entirely satisfactory in all chemical contexts.
However, we make the recommendation of
$\alpha=4$ based on the
analysis presented alongside Fig.~\ref{fig:RMSD_Metrics}.
%as it seems to be most widely successful
%across benchmark sets. %in the sense that
Taking a closer look, the \abwsaeq{4}\
error statistics suggest considerable improvements
relative to MP2 for NC, main-group TC (W4-11), TMTC, and ACONFL sets,
while minimal damage is done to the results for
H-atom\slash non-H-atom transfer barrier heights
and electronic properties.

While there is no universal parameter, \abws\
stands out from gap-dependent regularizers like
$\kappa$-MP2 (and the similarly-performing
$\sigma$-MP2 and $\sigma^2$-MP2 methods)\cite{SheRoiLet21}
in the sense that it is clearly more transferable
across different chemical problems.
This may be due to the fact that \abws\ defines a valid
second order BW correction for each $\alpha$.
As a consequence it incorporates
the full set of $\mathbf{t}$ amplitudes
in the regularizer, whereas gap-dependent
schemes rely only on the orbital
energy gaps.
The self-consistent nature of \abws\
may also act to further refine
the orbital energy gap, introducing
a feedback loop that fine-tunes
the resultant amplitudes.

As a final test for the robustness of our
parameterization, we consider
a secondary free parameter, $\beta$,
that directly modulates the amount of
\abws\ correlation energy such that,
$E = E_{\text{HF}} + \beta E_{\text{BW-s2($\alpha$)}}$.
The results in Section~\ref{sec:SI} show that the optimal $\beta$
parameter generally hovers in the range $0.9\leq\beta\leq1.1$.
Furthermore, when $\alpha$
nears its optimal value,
$\beta\rightarrow1.0$
with the exception of non-H-atom barrier heights
in NHTBH38 (Fig.~\ref{fig:TC_beta_alpha})
where $\beta=1.1$ when $\alpha=1$.
A $\beta>1$ implies systematic under-correlation,
and points to an optimal $\alpha$ for NHTBH38
that is less than 1.
In stark contrast to this, the landscape
of the parameter space for TMTC
in Fig.~\ref{fig:TMTC_beta_alpha}
features an optimal $\beta=0.7$ at low
$\alpha=1$, which increases to $\beta=1$ only when $\alpha\rightarrow8$.
This implies a significant over-correlation
for transition-metal systems that is
tempered only by larger $\alpha$ parameters.

The NC, W4-11, and ACONFL data sets in Figures~\ref{fig:NCI_beta_alpha}, \ref{fig:TC_beta_alpha},
and \ref{fig:ACONFL_beta_alpha}, respectively,
show a relatively flat slope
defined by the line tracing
$\min_{\alpha,\beta}\text{Error}(\alpha,\beta)$.
For these sets, the optimal $\beta$ is
very close to 1 across $\alpha$ parameters, suggesting that BW-s2($\alpha$) offers a
balanced description of correlation for NC,
main-group TC, and conformational isomers.
Overall, the relatively low slopes
across the parameter space
and the proximity of $\beta$ to 1
across various $\alpha$
both speak to the transferability
of the \abws\ approach.
Thus, moving forward we suggest
the single parameter \abwsaeq{4} approach
for general chemical applications.

\section*{Computational Details}
All calculations were performed in a development version
of Q-Chem~v6.0.2.\cite{QCHEM5}
All calculations (aside from evaluations of electronic properties)
feature SCF convergence thresholds that were set to $10^{-8}$
root-mean-square error. The correlation energy
was considered to be converged at a change of $10^{-8}$~Ha between iterations
for all calculations except for those
of the L7 dataset, where this was relaxed
to $10^{-5}$~Ha.
Relevant derivatives with respect to electric fields for
properties such as dipoles and polarizabilities
were evaluated via finite difference.
Because finite difference results are especially sensitive to
numerical errors, the SCF convergence and correlation
energy thresholds were set to $10^{-11}$.
To achieve complete basis set limit extrapolations
for NC, TC, and ACONFL we follow the protocol in
Ref.~\citen{NeeVal11}, which has been verified to perform
well with the heavy-aug-cc-pVDZ\slash heavy-aug-cc-pVTZ
basis sets used for L7.\cite{BalDunLao21}
For electronic response properties, we use the same
extrapolation method reported in Ref.~\citen{SheRoiLet21}.

We use restricted open-shell
orbitals which are separately pseudocanonicalized in the $\alpha$ and $\beta$ spaces before computing the correlation
energy in all open-shell systems.\cite{HubCar80,MurDav91,LauStaGau91,AmoAndNic91,KnoAndAmo91,LeeJay93}
For such systems, non-Brillouin singles (NBS) contributions
are included via,
\begin{equation}
	E_{\text{NBS}} = -\sum\limits_{ia}\frac{|F_{ia}|^2}{\varepsilon_a-\varepsilon_i}
\end{equation}
where $F_{ia}$ are off-diagonal Fock matrix elements.
%Hartree-Fock orbitals are used in all calculations.

Since $\mathbf{W}$ depends on the $\mathbf{t}$ amplitudes, which themselves
depend on the modulation of the energy gap supplied by the $\mathbf{W}$ matrix, the BW-s2 equations must be solved self-consistently.
We begin each BW-s2 calculation
with canonical Hartree-Fock orbitals
and an MP2 guess for the initial
$\mathbf{t}$ amplitudes, though
we note the possibility of obtaining
a strictly non-divergent initial guess
by means of Davidson's repartitioning
of the one-electron Fock
operator.\cite{Dav72}
To accelerate these calculations,
our implementation uses the resolution-of-the-identity (RI) approximation for the
two-electron integrals,\cite{FeyFitKom93,BerHar96}
resulting in a formal scaling of $m\times{\cal O}(N^5)$, where $m$ is the number of iterations
(typically between 4-6)
and $N$ is the number of basis functions.
Due to computational limitations,
the $I$ functions in the auxiliary RI basis sets were removed
for transition-metal calculations in the MOR39, MCO9, and
AuIrPt13 data sets.

\begin{acknowledgments}
This work was supported by the Director, Office of Science, Office of Basic Energy Sciences, of the U.S. Department of Energy under Contract No. DE-AC02-05CH11231. K. C.-F. acknowledges support from the National Institute Of General Medical Sciences of the National Institutes of Health under Award Number F32GM149165. The content is solely the responsibility of the authors and does not necessarily represent the official views of the National Institutes of Health.
\end{acknowledgments}

\section*{Author Declarations}
\subsection*{Conflict of Interest}
Martin Head-Gordon is a part-owner of Q-Chem, which is
the software platform used to perform the developments
and calculations described in this work.

\subsection*{Author Contributions}
\textbf{Kevin Carter-Fenk}: Writing -- original draft (equal); Writing -- review and editing (equal);
Investigation (lead);
Methodology (equal);
Software (lead); Data curation (lead).
\textbf{James Shee}: Writing -- original draft (equal);
Writing -- review and editing (equal);
Methodology (equal).
\textbf{Martin Head-Gordon}
Conceptualization (lead); funding
acquisition (lead); 
Writing -- review and editing (equal); Supervision (lead)

\section*{Data Availability}
Detailed data for NC, main-group TC, TMTC,
ACONFL, and electronic response properties
are available in the Supplementary Material.

\section*{References}
%\bibliography{allbib}


\makeatletter\@input{xx.tex}\makeatother

\end{document}

\title{
	Supporting Information for:\\
``Optimizing the Regularization in Size-Consistent Second-Order Brillouin-Wigner Perturbation Theory"
}
\author{Kevin Carter-Fenk$^1$,
        James Shee$^{1,2}$,
        and Martin Head-Gordon$^{1,3}$\\
$^1$\textit{Department of Chemistry, University of California, Berkeley, CA 94720, USA.}\\
$^2$\textit{Department of Chemistry, Rice University, Houston, TX 77005, USA}\\
$^3$\textit{Chemical Sciences Division, Lawrence Berkeley National Laboratory, Berkeley, CA 94720, USA}
}
\date{\today}
\maketitle
\clearpage\pagebreak

\section{Landscape of the $\alpha$\slash$\beta$ Parameter Space}\label{sec:SI}

\begin{figure}[h!!]
    \centering
    \fig{1.0}{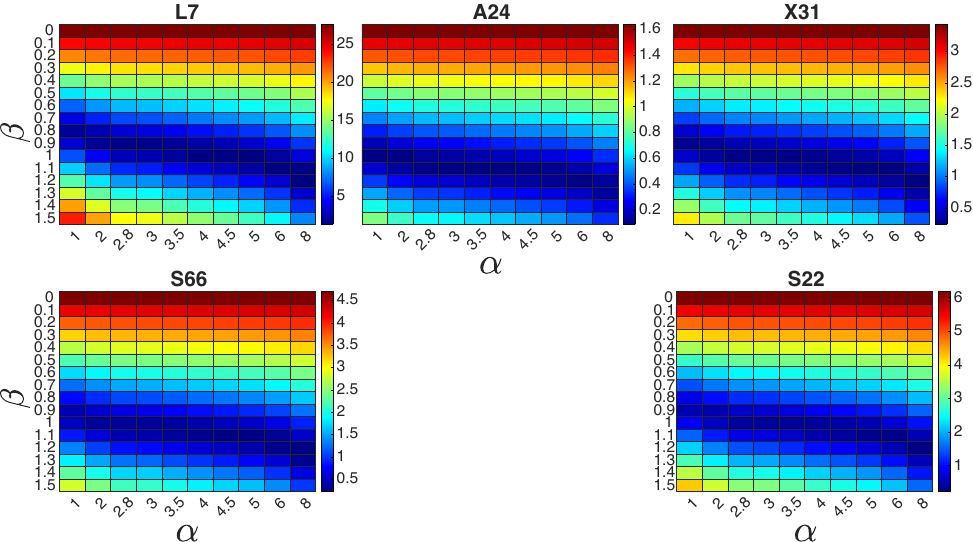}
    \caption{
    Scan across $\alpha$ and $\beta$ parameters
    for the NC data sets. Colors indicate
    the root-mean-square error (RMSE) for
    a given $\alpha$ and $\beta$ combination.
    }\label{fig:NCI_beta_alpha}
\end{figure}

\begin{figure}[h!!]
    \centering
    \fig{1.0}{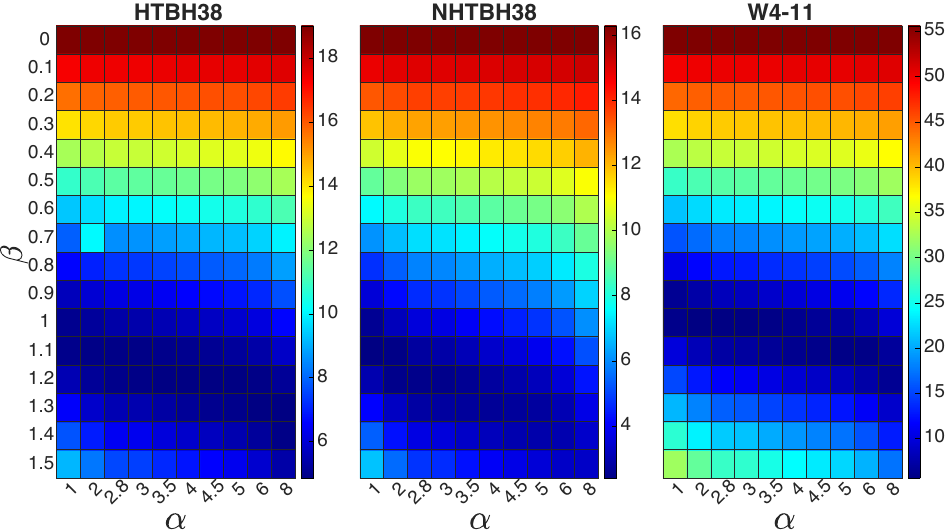}
    \caption{
    Scan across $\alpha$ and $\beta$ parameters
    for TC data sets. Colors indicate
    the root-mean-square error (RMSE) for
    a given $\alpha$ and $\beta$ combination.
    }\label{fig:TC_beta_alpha}
\end{figure}

\begin{figure}[h!!]
    \centering
    \fig{1.0}{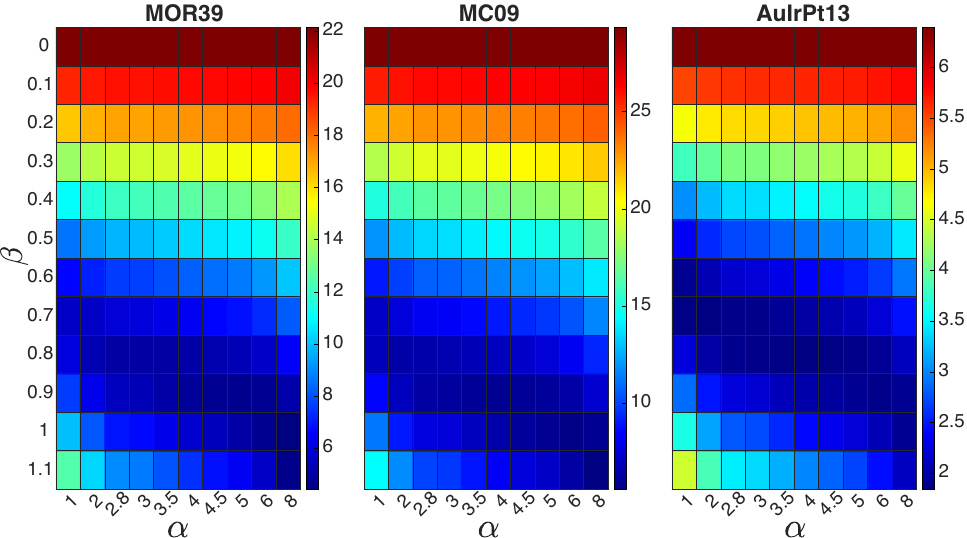}
    \caption{
    Scan across $\alpha$ and $\beta$ parameters
    for TMTC data sets. Colors indicate
    the root-mean-square error (RMSE) for
    a given $\alpha$ and $\beta$ combination.
    }\label{fig:TMTC_beta_alpha}
\end{figure}

\begin{figure}[h!!]
    \centering
    \fig{1.0}{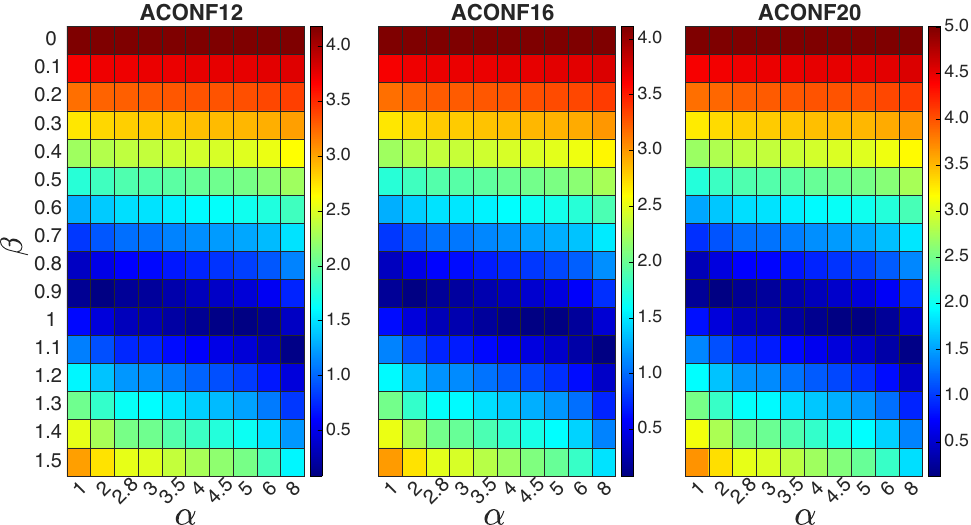}
    \caption{
    Scan across $\alpha$ and $\beta$ parameters
    for ACONFL data sets. Colors indicate
    the root-mean-square error (RMSE) for
    a given $\alpha$ and $\beta$ combination.
    }\label{fig:ACONFL_beta_alpha}
\end{figure}

\begin{figure}[h!!]
    \centering
    \fig{1.0}{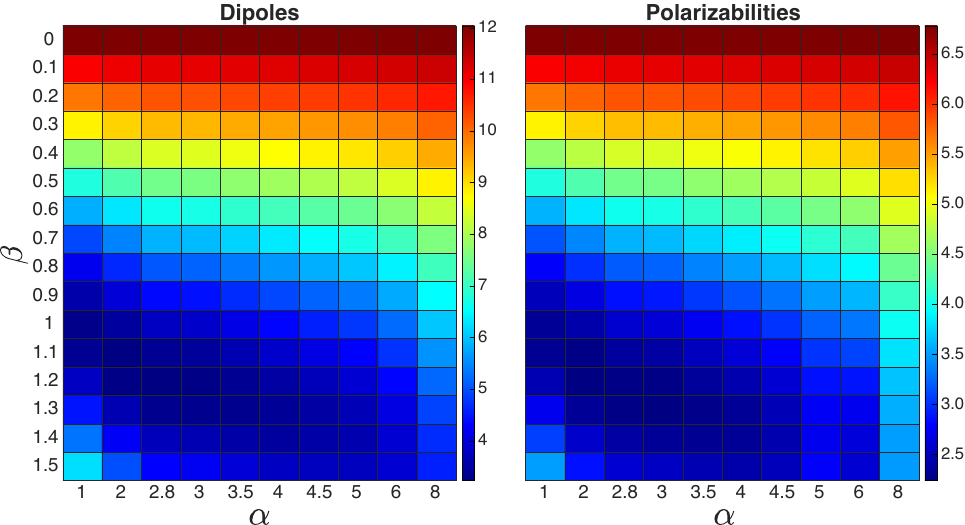}
    \caption{
    Scan across $\alpha$ and $\beta$ parameters
    for electronic properties data sets.
    Colors indicate
    the root-mean-square relative error (RMSRE) for
    a given $\alpha$ and $\beta$ combination.
    }\label{fig:ElectronicProperties_beta_alpha}
\end{figure}
\clearpage\pagebreak

\section{Graphical Representations of Error as a Function of $\alpha$}

\begin{figure}[h!!]
    \centering
    \fig{1.0}{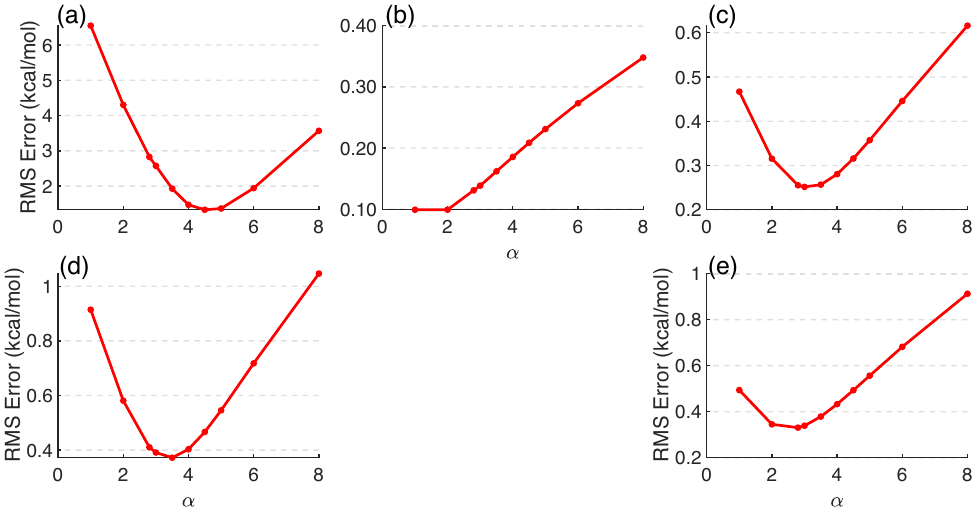}
    \caption{
    NC root-mean-square error
    cross section for $\beta=1$ of
    (a) L7, (b) A24, (c) X31, (d) S22,
    and (e) S66 data sets.
    }\label{fig:NCI_alphas}
\end{figure}

\begin{figure}[h!!]
    \centering
    \fig{1.0}{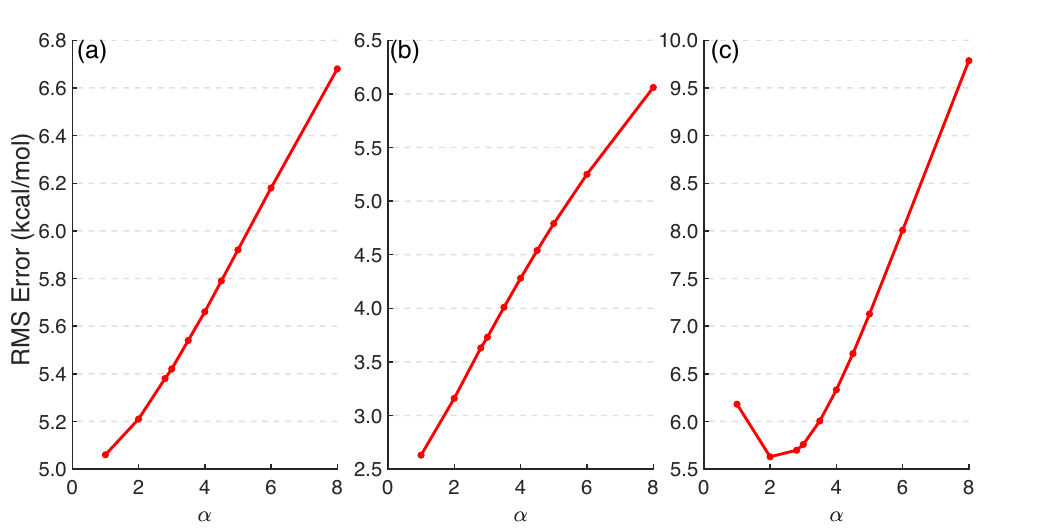}
    \caption{
    TC root-mean-square error
    cross section for $\beta=1$ of
    (a) HTBH38, (b) NHTBH38, and (c) W4-11 data sets.
    }\label{fig:TC_alphas}
\end{figure}

\begin{figure}[h!!]
    \centering
    \fig{1.0}{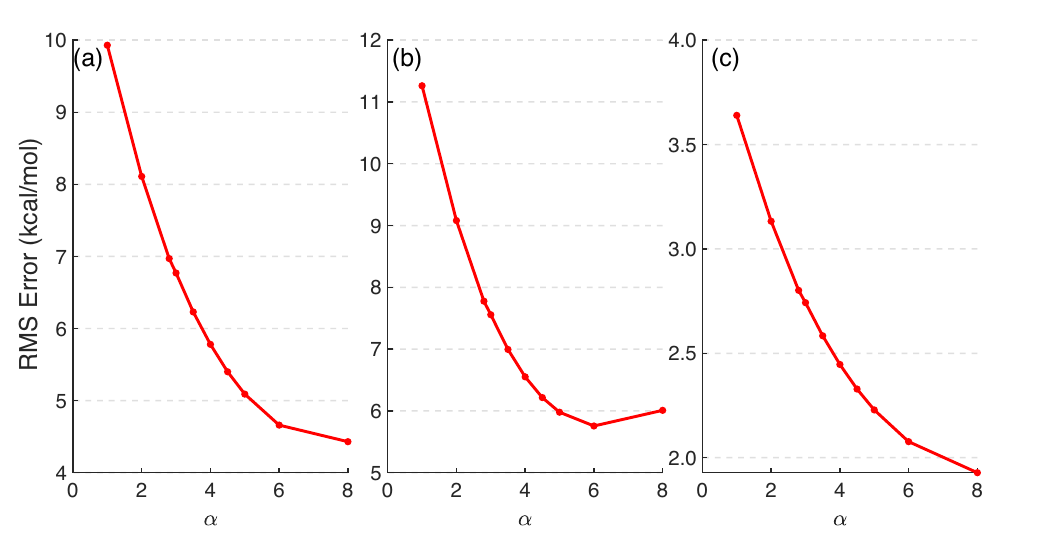}
    \caption{
    TMTC root-mean-square error
    cross section for $\beta=1$ of
    (a) MOR39, (b) MC09, and (c) AuIrPt13 data sets.
    }\label{fig:TMTC_alphas}
\end{figure}

\begin{figure}[h!!]
    \centering
    \fig{1.0}{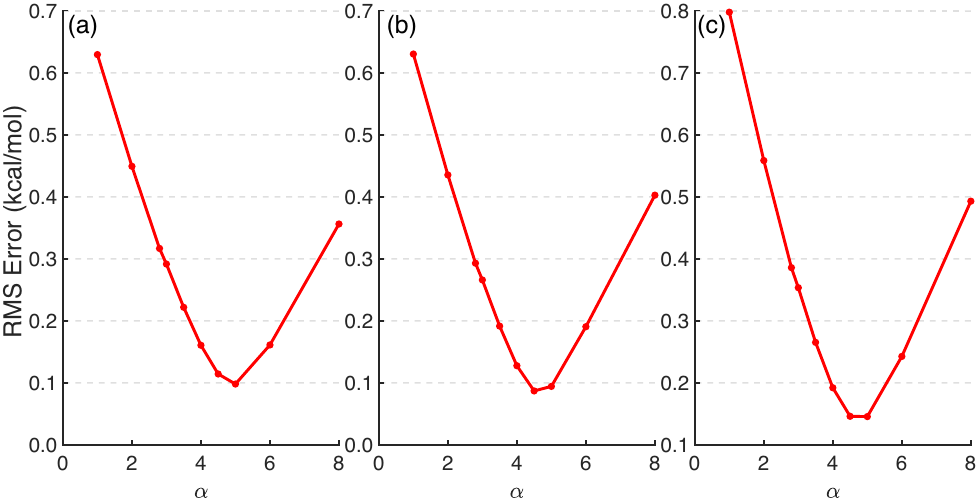}
    \caption{
    ACONFL root-mean-square error
    cross section for $\beta=1$ of
    (a) ACONF12, (b) ACONF16, and (c) ACONF20 data sets.
    }\label{fig:ACONFL_alphas}
\end{figure}

\begin{figure}[h!!]
    \centering
    \fig{1.0}{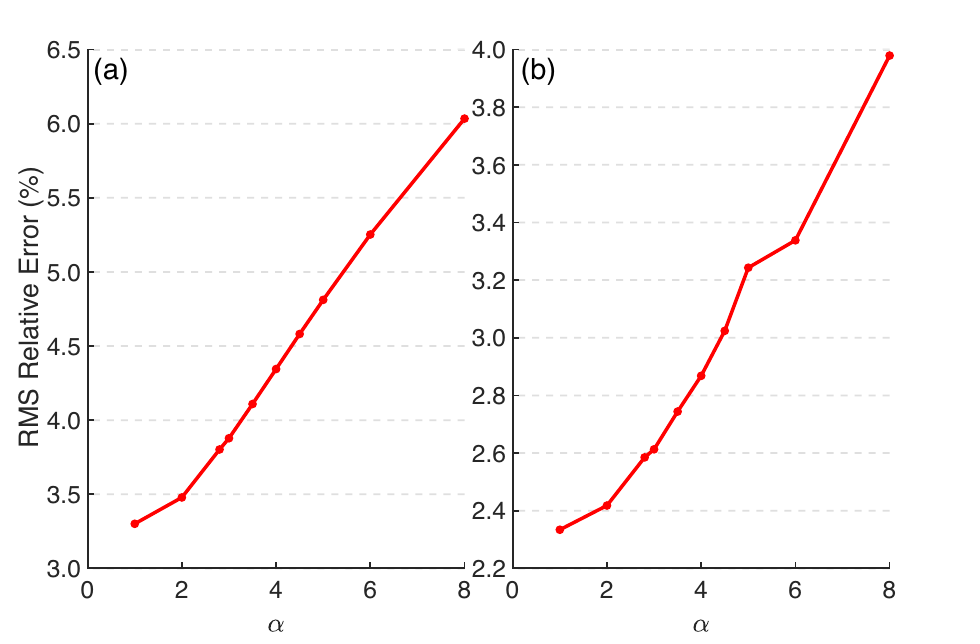}
    \caption{
    Electronic properties root-mean-square relative error cross section for $\beta=1$ of (a) dipole moments and (b) polarizabilities.
    }\label{fig:ElectronicProperties_alphas}
\end{figure}